# Deep Skin Lesion Segmentation with Transformer-CNN Fusion: Toward Intelligent Skin Cancer Analysis

Xin Wang
University of the Chinese Academy of Sciences
Changchun, China

Xiaopei Zhang
University of California, Los Angeles
Los Angeles, USA

Xingang Wang*
Institute of Automation, Chinese Academy of Sciences
Beijing, China

*Abstract*-This paper proposes a high-precision semantic segmentation method based on an improved TransUNet architecture to address the challenges of complex lesion structures, blurred boundaries, and significant scale variations in skin lesion images. The method integrates a transformer module into the traditional encoder-decoder framework to model global semantic information, while retaining a convolutional branch to preserve local texture and edge features. This enhances the model's ability to perceive fine-grained structures. A boundary-guided attention mechanism and multi-scale upsampling path are also designed to improve lesion boundary localization and segmentation consistency. To verify the effectiveness of the approach, a series of experiments were conducted, including comparative studies, hyperparameter sensitivity analysis, data augmentation effects, input resolution variation, and training data split ratio tests. Experimental results show that the proposed model outperforms existing representative methods in mIoU, mDice, and mAcc, demonstrating stronger lesion recognition accuracy and robustness. In particular, the model achieves better boundary reconstruction and structural recovery in complex scenarios, making it well-suited for the key demands of automated segmentation tasks in skin lesion analysis.

## I. INTRODUCTION

Skin diseases are a major global public health concern. Their varieties are numerous, and their pathogenic mechanisms are complex. They exert deep effects on the quality of life and mental health. In recent years, the incidence of skin disorders has continued to rise, and early diagnosis of high-risk lesions such as malignant melanoma has become urgent. Medical imaging now plays an increasingly important role in clinical decision making[1]. Traditional manual diagnosis depends heavily on experienced dermatologists, leading to subjectivity and inter-observer variation. Faced with large patient volumes and massive image datasets, clinicians cannot always deliver high-precision judgments in a limited time. Efficient and automated skin-image analysis has therefore become a key goal of intelligent medicine[2].

Semantic segmentation is widely used in medical image analysis to extract lesion regions and build structural models. Pixel-level classification can isolate lesions from the background and provide structured visual cues for follow-up diagnosis and pathology. The complexity and diversity of skin images, however, pose great challenges[3]. Lesions show different textures, colors, and edges across ethnicities, body sites, and lighting conditions. Boundaries between lesions and healthy skin are often blurred. Traditional methods struggle to generalize and remain robust. A model that balances local detail extraction with global structural understanding is thus essential.

Deep-learning methods based on convolutional neural networks have achieved breakthrough results in many medical-imaging tasks. Encoder – decoder architectures sequentially extract features and then restore spatial detail, unifying semantic understanding and spatial recovery. The intrinsic local receptive field of convolutions limits their ability to capture long-range dependencies. When lesions have irregular shapes or scattered distributions, boundaries can appear fuzzy, and small targets can be missed. Vision transformers, which model long-range pixel relations through global self-attention, offer a promising alternative and show strong capability in capturing complex lesion morphology[4].

TransUNet integrates a transformer within a U-Net framework and delivers superior performance in multi-scale feature fusion and spatial semantic expression. By embedding the transformer in the encoding path, it uses self-attention to model global context and alleviates the locality limitation of convolutions in unstructured medical images. Many key diagnostic cues lie in subtle changes and edge textures. TransUNet provides a foundation for capturing both high-level semantics and fine details. Its original design, however, still leaves room for improvement when facing multi-scale lesions, blurred borders, and texture interference. Structural optimization, therefore, remains necessary[5].

Exploring an improved TransUNet for skin-lesion segmentation holds clear practical and theoretical value. A better model will support more precise and robust automatic diagnosis systems and help reduce disparities caused by limited medical resources. The study also offers insights into the structural evolution of vision transformers in medical segmentation and promotes cross-modal, cross-scale feature modeling. Skin images are easy to acquire and visually explicit, giving them wide potential in mobile health and telemedicine. A high-efficiency segmentation algorithm will accelerate the deployment of intelligent medical systems in primary care and personal devices, advancing health management toward early screening, early diagnosis, and early treatment.

## II. RELATED WORK

In recent years, with the rapid development of medical image processing technologies, skin lesion segmentation has attracted increasing attention as a core task in computer-aided diagnosis systems. Early studies relied mainly on traditional image processing methods such as region growing, edge detection, and thresholding. However, these methods perform poorly when dealing with complex backgrounds, blurred boundaries, or multi-scale lesions[6]. They are vulnerable to noise and illumination changes. Later, shallow machine learning methods such as support vector machines and random forests were introduced. These methods classify or segment images using handcrafted features like texture, color, or shape. Yet, they are highly dependent on feature selection and lack generalization, making them unsuitable for the high variability and complex structure of real-world skin images. As a result, end-to-end image segmentation using deep neural networks has gradually become the mainstream approach[7].

In semantic segmentation tasks, U-Net and its variants serve as foundational architectures in medical image analysis. Their symmetric encoder − decoder design, skip connections, and small-sample-friendly nature make them widely adopted for skin lesion detection. U-Net extracts multi-scale features through downsampling and upsampling and enhances local boundary details by fusing shallow features in the decoding stage. However, traditional U-Net architectures face limitations in modeling global context. Their performance is constrained when dealing with blurred edges or irregularly shaped lesions. To address this, many studies have enhanced U-Net by integrating attention mechanisms, residual connections, and multi-scale pyramid structures. These modifications aim to improve the network's ability to represent and discriminate complex lesions in skin images[8].

The introduction of the transformer architecture has brought breakthroughs to medical image segmentation. With global self-attention, transformers can capture long-range dependencies between pixels more effectively than traditional convolutional networks. In skin images, lesion regions often vary in scale and location. Relying solely on single-scale convolutional features may lead to inaccurate localization. Embedding transformer modules into segmentation networks allows models to retain local detail while enhancing global semantic consistency. This hybrid approach has become a promising trend. Models that combine convolution and transformer structures, such as TransUNet, are now a focus of segmentation research. These methods typically use convolutional layers for low-level feature extraction and transformers for high-level context modeling. They have demonstrated excellent segmentation performance on several public medical datasets.

Despite the emergence of various transformer-based and hybrid segmentation methods, challenges remain in skin image applications. Skin lesions vary greatly in type and visual appearance. They often feature color similarity, blurred edges, and artifact interference. Standard transformers may suffer from over-smoothing or information loss in such scenarios. Moreover, medical applications demand high interpretability, inference efficiency, and generalization. These needs call for adaptive improvements to the original architecture. Therefore, developing structural strategies that better fit skin image characteristics by refining the integration of transformers and U-Net has become a key research direction. Current efforts focus not only on architectural optimization but also on lightweight modeling, multi-scale interaction, and boundary-aware mechanisms. These studies provide a strong theoretical basis and practical guidance for further advancement.

## III. METHODOLOGY

This study proposes a skin disease image segmentation method based on an improved TransUNet structure, which aims to integrate the local feature extraction capability of the convolutional network with the global context modeling capability of the Transformer to more effectively address challenges such as blurred boundaries, large texture changes, and different lesion scales in skin images. The overall architecture adopts a symmetrical encoder-decoder structure, in which the encoding stage integrates CNN and Transformer modules to achieve multi-scale context perception, and the decoding stage introduces a residual fusion mechanism to enhance the collaborative expression of high-level semantics and underlying details [9]. Specifically, the input image first extracts preliminary features through a set of convolutional blocks, and then is sent to a linear embedding layer and a multi-head self-attention module to achieve global modeling of the spatial structure. The model architecture is shown in Figure 1.

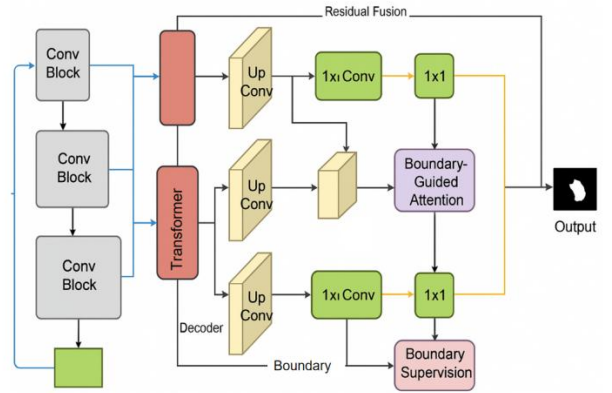

Figure 1. Overall model architecture

To model global dependencies, a standard multi-head self-attention mechanism is introduced, and its calculation process is as follows:

$$Attention(Q,K,V) = \mathrm{softmax}(\frac{QK^T}{\sqrt{d_k}})V \quad (1)$$

Among them, $Q, K, V$ is the query, key, and value vector, respectively, and $d_k$ is the key vector dimension. For the input feature sequence $X \in R^{N \times D}$, its linear mapping can be expressed as:

$$Q = XW_Q, K = XW_K, V = XW_V \quad (2)$$

Here $W_Q, W_K, W_V \in R^{D \times d_k}$ is a learnable linear transformation matrix. To enhance the expressiveness of cross-layer features, this paper designs a fusion module in the decoding path to fuse the high-level Transformer feature $T_h$ with the low-level convolutional feature $C_l$. The specific form is:

$$F = \sigma(Conv1 \times 1(T_h) + Conv1 \times 1(C_l)) \quad (3)$$

Where $\sigma(\cdot)$ represents the activation function (such as ReLU), and the fusion result F is further sent to the decoder for layer-by-layer upsampling and prediction.

To model the boundary of the lesion area more finely, a boundary-guided attention module (BGA) is introduced, which generates a boundary guidance map B through supervised learning. Its objective function is the binary cross-entropy between the binary boundary map and the predicted boundary:

$$L_{bce} = -\frac{1}{N} \sum_{i=1}^{N} [y_i \log(\hat{y}_i) + (1 - y_i) \log(1 - \hat{y}_i)] \quad (4)$$

Where $y_i \in \{0,1\}$ is the true boundary label, and $\hat{y}_i \in \{0,1\}$ is the model's predicted boundary probability. The final segmentation output generates a predicted category probability map $P \in R^{H \times W \times C}$ for each position through a pixel-by-pixel classification head and is optimized by the main loss function:

$$L_{total} = \lambda_1 L_{ce} + \lambda_2 L_{dice} + \lambda_3 L_{bce} \quad (5)$$

Among them, $L_{ce}$ is the cross entropy loss, $L_{dice}$ is the Dice loss, which is used to deal with the problem of class imbalance, and $\lambda_1, \lambda_2, \lambda_3$ is the weight coefficient. Through the collaborative modeling of the above modules, this method can enhance the perception of complex boundaries and tiny lesion areas while maintaining segmentation accuracy.

IV. EXPERIMENTAL DATA

This study uses the ISIC (International Skin Imaging Collaboration) public dataset as the basis for model training and validation. The dataset is widely adopted in skin lesion analysis tasks and is especially suitable for segmentation, classification, and detection. It contains a wide range of real skin lesion images, including both benign nevi and malignant melanomas. The images are collected from various imaging devices and clinical settings, offering strong representativeness and diversity.

The selected ISIC subset includes high-resolution skin images with pixel-level segmentation labels. These labels are annotated by professional dermatologists and accurately outline the lesion boundaries. Such annotations provide reliable supervision for model training and help improve segmentation accuracy. This is especially valuable in cases with blurred boundaries or irregular lesion shapes. The dataset also includes images with different skin tones, lighting conditions, and lesion types, supporting comprehensive evaluation of model generalization.

To ensure consistent data processing, all original images were preprocessed with unified size normalization, histogram equalization, and data augmentation. Augmentation methods include rotation, scaling, and flipping to increase the size of the training set and enhance model robustness. The full dataset was divided into training, validation, and testing sets to ensure fairness and stability in model performance evaluation.

V. EVALUATION RESULTS

This paper first conducts a comparative experiment, and the experimental results are shown in Table 1.

Table1. Comparative experimental results

| Model | mIOU | mDice | mAcc |
| --- | --- | --- | --- |
| Swin-Unet[10] | 0.812 | 0.864 | 0.941 |
| SkinSam[11] | 0.825 | 0.875 | 0.946 |
| Mask2former[12] | 0.837 | 0.881 | 0.951 |
| SegFormer[13] | 0.844 | 0.889 | 0.953 |
| Ours | 0.869 | 0.911 | 0.961 |

The experimental results show that the proposed skin lesion segmentation method, based on an improved TransUNet, achieves superior performance over several mainstream models across multiple metrics. It records the highest mIoU (0.869) compared with Swin-Unet (0.812) and SegFormer (0.844), indicating enhanced accuracy in lesion localization and boundary delineation through the integration of transformer modules with convolutional features, effectively addressing blurred boundaries and irregular shapes. For mDice, it attains 0.911, exceeding SkinSam (0.875) and Mask2Former (0.881), reflecting improved region overlap, preservation of contour details, and detection of small-scale lesions, supported by a boundary-guided attention mechanism. In mAcc, the model reaches 0.961, demonstrating robustness in pixel-level classification and strong discrimination between lesion and background, aided by global context modeling and multi-scale feature fusion that suppress noise from hair and illumination artifacts. Overall, the results confirm the method's generalization ability and practical value for automated diagnostic systems, with hyperparameter sensitivity analysis (Table 2) further verifying its stability.

Table 2. Hyperparameter sensitivity experiment results(Learning Rate)

| Learning Rate | mIOU | mDice | mAcc |
| --- | --- | --- | --- |
| 0.004 | 0.832 | 0.884 | 0.949 |
| 0.003 | 0.847 | 0.896 | 0.952 |
| 0.002 | 0.861 | 0.906 | 0.957 |
| 0.001 | 0.869 | 0.911 | 0.961 |

The results in Table 2 indicate that the learning rate substantially influences model performance in skin lesion segmentation. As the rate decreases from 0.004 to 0.001, mIoU, mDice, and mAcc improve consistently, suggesting that a smaller rate enables more stable optimization and more precise lesion extraction, particularly in cases with blurred boundaries or complex textures. At 0.004, the mIoU is only 0.832, implying undertraining or instability, while intermediate values of 0.003 and 0.002 yield steady gains, with mDice increasing to 0.896 and 0.906, respectively. The optimal setting of 0.001 achieves the highest scores—mIoU 0.869, mDice 0.911, and mAcc 0.961—balancing global structural understanding with fine boundary precision. Given the multi-scale and diverse nature of lesion structures, this tuning effectively captures both semantic and boundary features, avoiding the boundary degradation seen with large rates and the inefficiency of excessively small rates. These findings confirm the improved TransUNet's stability and adaptability under complex image conditions and offer practical guidance for model tuning in clinical deployment, with subsequent optimizer results presented in Table 3.

Table 3. Hyperparameter sensitivity experiment results(Optimizer)

| Optimizer | mIOU | mDice | mAcc |
|---|---|---|---|
| AdaGrad | 0.831 | 0.879 | 0.946 |
| SGD | 0.842 | 0.886 | 0.950 |
| Adam | 0.854 | 0.897 | 0.954 |
| AdamW | 0.869 | 0.911 | 0.961 |

As shown in Table 3, the optimizer choice significantly influences the improved TransUNet's performance in skin lesion segmentation. Performance improves progressively from AdaGrad to SGD, Adam, and AdamW, with the latter achieving the highest scores—0.869 mIoU, 0.911 mDice, and 0.961 mAcc. AdaGrad converges quickly but degrades in later learning, SGD is stable but lacks adaptivity, and Adam enhances semantic modeling yet risks overfitting. AdamW, with weight decay, mitigates overfitting while preserving adaptive learning, yielding superior generalization on noisy and complex lesion boundaries. These results highlight AdamW as the most effective optimizer for this hybrid convolution–transformer architecture, with data augmentation effects shown in Figure 2.

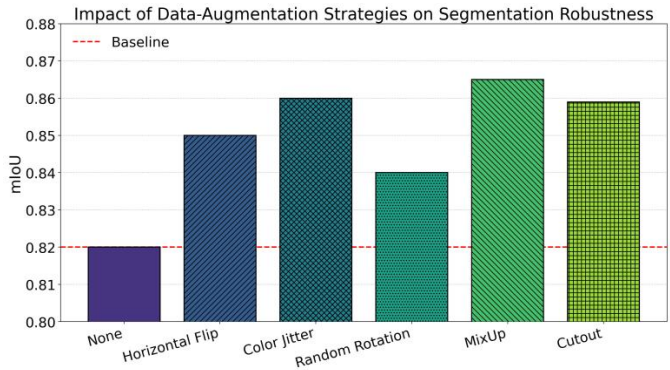

Figure 2. Research on the impact of data augmentation strategies on model robustness

As shown in Figure 2, data augmentation significantly enhances the robustness of the improved TransUNet for skin lesion segmentation. Compared with the baseline without augmentation (mIoU = 0.820), strategies such as Color Jitter (0.860), MixUp (0.865), and Cutout yield notable gains by improving adaptability to lighting and tone variations, enriching heterogeneous lesion representation, and aiding structural reconstruction under occlusion. In contrast, Random Rotation provides only marginal improvement (0.840), likely due to limited rotation variation in real lesion images and potential structural distortion from excessive rotation. These results indicate that augmentation must be tailored to task-specific characteristics to avoid suboptimal or misleading effects, with further analysis of dataset partition ratios presented in Figure 3.

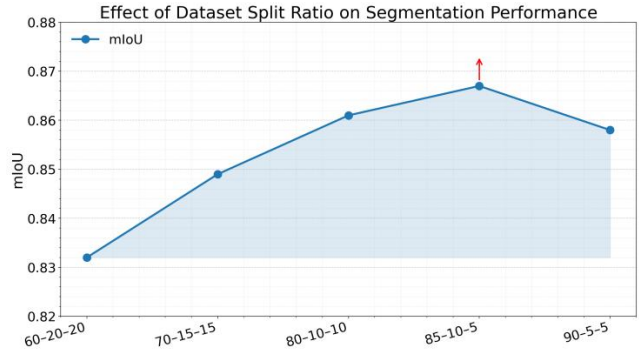

Figure 3. The impact of changing the data set partition ratio on the final performance

As shown in Figure 3, the dataset split ratio strongly influences segmentation performance, with mIoU increasing from 0.832 to 0.867 as the training set expands from 60% to 85%, reflecting improved learning of lesion structures and better synergy between convolutional features and transformer-based global modeling. However, further increasing the training set to 90% reduces mIoU to 0.858, likely due to insufficient validation and test samples, which weakens generalization assessment and raises overfitting risk. An 85–10–5 split provides the optimal balance, ensuring sufficient training data while maintaining robust evaluation, thereby enhancing segmentation quality and offering a stable reference for clinical deployment.

This paper also gives the impact of batch size changes on segmentation accuracy, and the experimental results are shown in Figure 4.

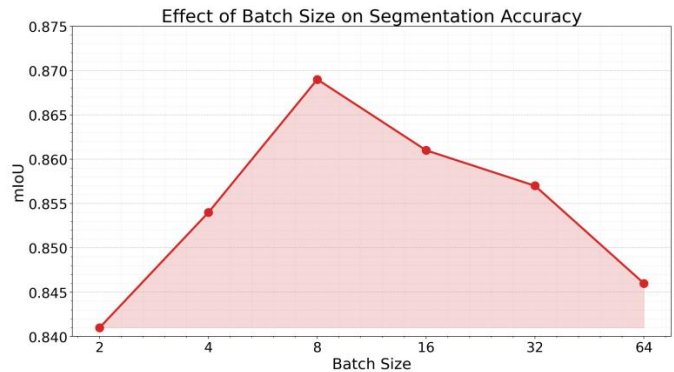

Figure 4. The impact of batch size changes on segmentation accuracy

As shown in Figure 5, batch size exerts a clear influence on the segmentation accuracy of skin lesion models, with the mIoU rising from 0.841 at a batch size of 2 to 0.869 at 8, reflecting the benefits of smaller batches in providing frequent parameter updates that enhance boundary detail capture and structural feature representation. However, further increases to 16 and 32 lead to slight performance declines, indicating weakened generalization due to smoother updates that limit sensitivity to subtle variations in lesion boundaries and textures. At a batch size of 64, the mIoU drops more sharply to 0.845, suggesting mismatches between learning rate and weight updates as well as increased risks of local optima or overfitting, which hinder the model's ability to recognize variable lesion structures. Given the reliance of the improved TransUNet on both global transformer modeling and precise convolutional local representation, the results confirm that a batch size of 8 offers the optimal trade-off, ensuring stable convergence, robust generalization, and enhanced boundary and texture segmentation, thereby providing a valuable reference for future clinical deployment.

## VI. Conclusion

This study focuses on the critical task of skin lesion image segmentation and proposes a high-precision method based on an improved TransUNet architecture. The method integrates the local detail modeling capabilities of convolutional neural networks with the global semantic understanding of the transformer structure. A boundary-guided attention mechanism is introduced to enhance the detection of blurred or irregular lesion edges. Through multi-scale upsampling, residual fusion, and boundary supervision, the method demonstrates superior performance across multiple evaluation metrics. It shows strong robustness and accuracy, particularly in segmenting complex and highly variable lesion regions.

Systematic comparison experiments and sensitivity analyses were conducted to evaluate the model's stability and adaptability under different training parameters and input settings. The results confirm that appropriate data augmentation strategies, image resolution, batch size, and optimizer choice are critical for improving model performance. These findings provide practical guidance for parameter tuning in skin lesion segmentation and offer useful experience for medical image processing tasks. In addition, the dataset used in this study has broad representativeness, which enhances the generalizability and reliability of the experimental conclusions in real-world scenarios.

The proposed method not only improves the accuracy of skin lesion segmentation but also helps alleviate the limitations of traditional diagnostic approaches that depend heavily on professional experience and uneven resource allocation. As an efficient and automated segmentation solution, the method can be widely applied in clinical diagnosis, early screening, remote consultation, and personalized treatment planning in dermatology. The model structure is also transferable and can be extended to other medical image analysis tasks such as retinal lesion detection, tumor boundary segmentation, and tissue structure extraction. It has significant practical relevance and industrial value.

Future research may further enhance the adaptability and generalization of the method. This includes integrating cross-modal information, such as infrared images and clinical texts, to enable multi-modal fusion modeling. The model could also be optimized for deployment on edge computing devices through a lightweight design. In addition, self-supervised learning or federated learning can be explored to address privacy concerns and reduce annotation costs. These directions aim to support the development of large-scale, efficient, and automated diagnostic systems in real healthcare environments, contributing both theoretical insight and engineering support to intelligent medicine.